\documentclass[showpacs,aps,graphicx,twocolumn]{revtex4}
\usepackage{amsmath}
\usepackage{graphicx}

\begin{document}

\title{One-step error correction  for  multipartite polarization entanglement\footnote{Published in Phys. Rev. A \textbf{83}, 062316 (2011)}}

\author{Fu-Guo Deng\footnote{Email:
fgdeng@bnu.edu.cn} }
\address{  Department of Physics, Applied Optics Beijing Area Major
Laboratory, Beijing Normal University, Beijing 100875, China }
\date{\today }

\begin{abstract}
We present two economical one-step error-correction protocols for
multipartite polarization-entangled systems in a
Greenberger-Horne-Zeilinger state. One uses spatial entanglement to
correct errors in the polarization entanglement of an N-photon
system, resorting to linear optical elements. The other uses
frequency entanglement to correct errors in the polarization
entanglement of an N-photon system. The parties in quantum
communication can obtain a maximally entangled state from each
N-photon system transmitted with one step in these two protocols,
and both of their success probabilities are 100\%, in principle.
That is, they both work in a deterministic way, and they do not
largely consume the less-entangled photon systems, which is far
different from conventional multipartite entanglement purification
schemes. These features may make these two protocols more useful for
practical applications in long-distance quantum communication.
\end{abstract}
 \maketitle

\section{introduction}

Entanglement is an important quantum resource for quantum
information processing \cite{book}. The powerful speedup of quantum
computation resorts to multipartite entanglement. Long-distance
quantum communication should resort to entanglement for setting up
the quantum channel between  remote locations, including quantum
teleportation \cite{teleportation}, quantum key distribution (QKD)
\cite{Ekert91,BBM92,CORE,LongLiu,rmp}, quantum dense coding
\cite{densecoding,densecoding2,densecoding3}, quantum secret sharing
\cite{QSS,QSS2,QSS3}, and so on. In a practical transmission, an
entangled quantum system will inevitably interact with its
environment, which will degrade the entanglement of the system. In
general, the interaction will make an entangled system be in a mixed
state. The decoherence of entanglement in quantum system will affect
quantum communication largely. For example, it will decrease the
security of a QKD protocol if a maximally entangled state
transmitted  over a noisy channel becomes a mixed entangled state as
a vicious eavesdropper can exploit the decoherence to hide her
illegal action. The non-maximally entangled quantum channel will
decrease the fidelity of quantum teleportation and quantum dense
coding.

Entanglement purification
\cite{Bennett1996,Deutsch,Pan1,Simon,shengpra2008,wangcpra,shengpratwostep,shengpraonestep,lixhepp}
is an interesting tool for the parties in quantum communication to
improve the fidelity of entangled quantum systems after they are
transmitted over a practical channel. Its task is to obtain a subset
of high-fidelity entangled quantum systems from a set of quantum
systems in a mixed entangled state. In 1996, Bennett \emph{et al.}
\cite{Bennett1996} proposed an original entanglement purification
protocol (EPP) to purify a Werner state \cite{werner}, resorting to
quantum controlled-not (CNOT) gates and bidirectional unitary
operations. Subsequently, Deutsh \emph{et al}. \cite{Deutsch}
improved this EPP with  two additional specific unitary operations.
In 2001, Pan \emph{et al}. \cite{Pan1} presented an EPP with linear
optical elements and an ideal entanglement source by sacrificing a
half of the efficiency in the EPP by Bennett \emph{et al.} However,
this protocol decreases largely the difficulty for the
implementation of EPP with current technology as it is impossible to
construct a perfect CNOT gate based on linear optics. In 2002, Simon
and Pan \cite{Simon} proposed an EPP with a  currently available
parametric down-conversion (PDC) source, not an ideal single-pair
entanglement source. This protocol exploits the spatial entanglement
of a photon pair to purify its polarization entanglement. In 2003,
Pan \emph{et al.} \cite{Pan2} demonstrated this EPP in experiment.
In 2008, an efficient EPP \cite{shengpra2008} based on a PDC source
was proposed with cross-Kerr nonlinearity. It has the same
efficiency as the EPP by Bennett \emph{et al.} with perfect CNOT
gates. Moreover, it can be repeated to get a high-fidelity entangled
photon pairs from a practical entanglement source. However, the
cross-Kerr nonlinearity is too small in nature, which increases the
difficulty for the implementation of this EPP at present. In 2008,
Xiao \emph{et al.} \cite{wangcpra} proposed an EPP with frequency
entanglement.

Recently, the concept of deterministic entanglement purification was
proposed \cite{shengpratwostep} for two-photon entangled systems,
which is far different from the conventional entanglement
purification (CEPP)
\cite{Bennett1996,Deutsch,Pan1,Simon,shengpra2008,wangcpra} as the
former works in a deterministic way, while the latter works in a
probabilistic way. In 2010, we introduced a two-step deterministic
entanglement purification protocol (DEPP) \cite{shengpratwostep} for
entangled photon pairs, resorting to hyerentanglement. In this
two-step DEPP, the spatial entanglement and  the frequency
entanglement of a quantum system are used to correct the bit-flip
error and  the phase-flip error in this quantum system,
respectively. With two error-correction processes, the two parties
in quantum communication, say Alice and Bob, can obtain a maximally
entangled state from each entangled system in theory. Certainly, it
is difficult to implement this two-step DEPP in experiment at
present as it requires the hyperentanglement of photon systems and a
nonlinear optical interaction. Subsequently, we proposed an one-step
DEPP \cite{shengpraonestep}, resorting to  the spatial entanglement
of a practical PDC source and linear optical elements. With this
one-step DEPP, Alice and Bob can in principle obtain a maximally
entangled photon pair from each photon system with only one step.
Simultaneously, Li \cite{lixhepp} presented independently an
interesting DEPP using spatial entanglement, called it also an
one-step DEPP as it has the same role as the one-step DEPP in Ref.
\cite{shengpraonestep} although there are, in essence, some
differences in explaining the principle of the detereministic
entanglement purification. Although the physics in the one-step DEPP
in Ref. \cite{shengpraonestep} is more clear than that in Ref.
\cite{lixhepp}, the one-step DEPP by Li is more economical than that
in Ref. \cite{shengpraonestep} as the former needs two polarizing
beam splitters (PBSs), while the latter needs eight PBSs.

In essence, each of all existing EPPs
\cite{Bennett1996,Deutsch,Pan1,Simon,shengpra2008,wangcpra,shengpratwostep,shengpraonestep,lixhepp}
can be considered as a quantum error-correction protocol. In CEPPs,
the process for error correction is divided into two steps which are
repeated some times for improving the fidelity of entanglement
largely. One step is used to correct the bit-flip error and the
other is used correct the phase-flip error. That is, all the CEPPs
\cite{Bennett1996,Deutsch,Pan1,Simon,shengpra2008,wangcpra} work in
a probabilistic way for the correction of the errors in only one
degree of freedom of photons. In DEPPs
\cite{shengpratwostep,shengpraonestep,lixhepp}, the error correction
is completed with one step. That is, they work in a deterministic
way, but they should resort to at least another degree of freedom of
photons.

By far, there have been several interesting EPPs
\cite{Bennett1996,Deutsch,Pan1,Simon,shengpra2008,shengpratwostep,shengpraonestep,lixhepp,wangcpra}
focusing on the bipartite entangled quantum systems, while the
number of multipartite entanglement purification protocols (MEPP)
\cite{Murao,shengepjd,shengpla} is very small as the structure of a
multipartite quantum system is more complicated than that of a
bipartite quantum system. Moreover, it is difficult to optimize a
MEPP as its efficiency depends on a great number of parameters
coming from the channel noise. In 1998, Murao \emph{et al.}
\cite{Murao} proposed the first multipartite entanglement
purification protocol (MEPP) to purify multipartite quantum system
in a Greenberger-Horne-Zeilinger (GHZ) with CNOT gates, following
some ideas in the EPP by Bennett et al \cite{Bennett1996}. In 2009,
a MEPP based on cross-Kerr nonlinearities was proposed
\cite{shengepjd}. In this protocol, the cross-Kerr nonlinearity is
used to construct a nondestructive quantum nondemolition detector
(QND) \cite{QND1} which has the functions of both a parity-check
detector and a single-photon detector. With QNDs, the parties can
obtain some high-fidelity GHZ states from an ensemble in a mixed
entangled state by performing this MEPP repeatedly. In 2009, a
multipartite electronic entanglement purification was proposed with
charge detection \cite{shengpla}. All these three MEPPs work in a
conventional way. That is, the parties can only obtain a subset of
high-fidelity entangled states from an ensemble in a mixed
less-entangled state by performing  the MEPPs repeatedly and
sacrificing a great deal of quantum resource.

In this paper, we will present two economical one-step
error-correction  protocols for multipartite polarization-entangled
systems in a GHZ state. The first one-step multipartite polarization
entanglement error-correction protocol (MPEECP) is based on spatial
entanglement and simple linear optical elements. The parties in
quantum communication can obtain a maximally entangled state from
each system polluted by the channel noise on the polarization degree
of freedom. Also, the number of the linear optical elements in this
protocol is reduced to be a minimal one. Moreover, this protocol
works in a deterministic way, which is far different from the
polarization entanglement purification protocol for two-photon
system by using the spatial entanglement \cite{Simon} as the latter
is used to improve the fidelity of an ensemble in a mixed entangled
state by repeating the protocol again and again. The second MPEECP
uses the frequency entanglement to correct the errors in the
polarization entanglement of an $N$-photon system and reduces the
number of the channels for the transmission of the system. The
success probabilities of these two economical one-step MPEECPs are
in principle 100\%.  Compared with the conventional multipartite
entanglement purification protocols \cite{Murao,shengepjd,shengpla},
these two  economical one-step MPEECPs reduces the quantum resource
consumed largely. These advantages maybe make these two MPEECPs more
useful in practical applications in long-distance quantum
communication in future.

\section{Economic one-step MPEECP with spatial entanglement}

As shown by Simon and Pan \cite{Simon}, an entangled state of a
photon pair in two degrees of freedom can be written as
\begin{eqnarray}
|\Psi\rangle=\frac{1}{2}(|a_{1}\rangle|b_{1}\rangle+|a_{2}\rangle|b_{2}\rangle)(|H\rangle|H\rangle
+|V\rangle|V\rangle)_{ab},\label{state1}
\end{eqnarray}
where $|H\rangle$ and $|V\rangle$ represent the horizonal and the
vertical polarizations of photons, respectively.   The subscripts
$a$ and $b$ represent the two photons sent to the two parties, say
Alice and Bob, respectively.  $a_1$ ($b_1$) and $a_2$ ($b_2$) are
the two spatial modes for the photon sent to Alice (Bob). As the
spatial entanglement is far more stable than the polarization
entanglement over an optical-fiber channel, Simon and Pan
\cite{Simon} exploited the spatial entanglement of a photon pair
coming from a PDC source to purify the polarization of a photon
pair. By controlling the phase stability, the EPP by Simon and Pan
can improve the fidelity of an ensemble in a mixed entangled state,
as shown in the experiment by Pan \emph{et al.} \cite{Pan2}.

The GHZ state of a multipartite entangled system composed of $N$
two-level particles can be described as
\begin{eqnarray}
|\Phi^{+}\rangle_N=\frac{1}{\sqrt{2}}(|HH\cdots H\rangle + |VV\cdots
V\rangle)_{AB\cdots C}. \label{state2}
\end{eqnarray}
Here the subscripts $A$, $B$, $\cdots$, and  $C$ represent the
photons sent to the parties Alice, Bob, $\cdots$, and Charlie,
respectively. Certainly, there are other $2^N-1$ GHZ state for an
$N$-qubit system and can be written as $|\Phi^{+}\rangle_{ij\cdots
k}=\frac{1}{\sqrt{2}}(|ij\cdots k\rangle + |\bar{i}\bar{j}\cdots
\bar{k}\rangle)_{AB\cdots C}$ and
$|\Phi^{-}\rangle_{\bar{i}\bar{j}\cdots
\bar{k}}=\frac{1}{\sqrt{2}}(|ij\cdots k\rangle -
|\bar{i}\bar{j}\cdots \bar{k}\rangle)_{AB\cdots C}$ . Here
$i,j,k\in\{0,1\}$, $\bar{i}=1-i$, $\bar{j}=1-j$, and $\bar{k}=1-k$.
$\vert 0\rangle\equiv \vert H\rangle$ and $\vert 1\rangle\equiv
\vert V\rangle$. An entangled state of an $N$-photon system in the
polarization and the spatial-mode degrees of freedom can be written
as
\begin{eqnarray}
|\Phi\rangle_s &=& \frac{1}{2}(|a_1\rangle|b_1\rangle \cdots \vert
c_1 \rangle + |a_2\rangle|b_2\rangle \cdots \vert c_2
\rangle)\cdot\nonumber\\
&& (|H\rangle \vert H\rangle\cdots|H\rangle +|V\rangle\vert
V\rangle\cdots |V\rangle)_{AB\cdots C}.\label{state11}
\end{eqnarray}
Here  $\vert \phi\rangle_s=\frac{1}{\sqrt{2}}(|a_1\rangle|b_1\rangle
\cdots \vert c_1 \rangle + |a_2\rangle|b_2\rangle \cdots \vert c_2
\rangle)$ represents the spatial entanglement of an $N$-photon
system.

\begin{widetext}
\begin{center}
\begin{figure}[!h]
\begin{center}
\includegraphics[width=12cm,angle=0]{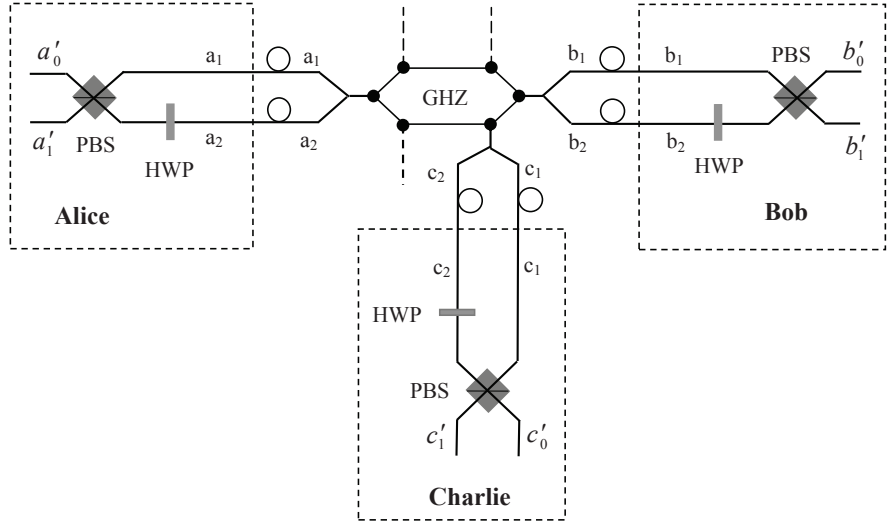}
\caption{Schematic illustration for the principle of the present
economic one-step  multipartite polarization entanglement
error-correction protocol using spatial entanglement. PBS represents
a polarizing beam splitter and it is used to transfer a $|H\rangle$
polarization photon and reflet a  $|V\rangle$ polarization photon.
HWP represents a half wave plate and it can convert  $|H\rangle$
into $|V\rangle$, and $|V\rangle$ into $|H\rangle$. After the
detection for each spatial mode, the parties in quantum
communication can obtain a maximally entangled polarization state
from each multi-photon system in a deterministic way in principle.
}\label{fig2multiparticlespatial}
\end{center}
\end{figure}
\end{center}
\end{widetext}

The principle of the present economic one-step MPEECP using spatial
entanglement is shown in Fig. \ref{fig2multiparticlespatial}.
Suppose that the original state of an $N$-photon system is $\vert
\Phi\rangle_s$. After the photons suffer from the noise in the
channels, the state of the system becomes
\begin{eqnarray}
\rho=\rho_p\cdot \rho_s,\label{state41}
\end{eqnarray}
where $\rho_p$ and $\rho_s$ represent the states of the multi-photon
systems in the polarization and the spatial degrees of freedom,
respectively, i.e.,
\begin{eqnarray}
\rho_s=\vert \phi\rangle_s\langle \phi\vert, \label{state51}
\end{eqnarray}
\begin{eqnarray}
\rho_p=\sum_{i,j,\cdots,k}F_{ij\cdots k}\vert
\Phi^+\rangle_{ij\cdots k}\langle \Phi^+\vert +
F_{\bar{i}\bar{j}\cdots \bar{k}}\vert \Phi^-\rangle_{ij\cdots
k}\langle \Phi^-\vert. \label{state62}
\end{eqnarray}
\begin{eqnarray}
\sum_{i,j,\cdots,k}F_{ij\cdots k} + F_{\bar{i}\bar{j}\cdots
\bar{k}}=1. \label{state71}
\end{eqnarray}
Here $F_{ij\cdots k}$ ($F_{\bar{i}\bar{j}\cdots \bar{k}}$) is the
probability that the $N$-photon system is in the state $\vert
\Phi^+\rangle_{ij\cdots k}$ ($\vert \Phi^-\rangle_{ij\cdots k}$)
after it is transmitted over a noisy channel. That is, the mixed
state shown in Eq.(\ref{state41}) can be viewed as a probabilistic
mixture of $2^N$ pure states. The system is in the state $\vert
\Phi^+\rangle_{ij\cdots k}\cdot \vert \phi\rangle_s$ or the state
$\vert \Phi^-\rangle_{ij\cdots k}\cdot \vert \phi\rangle_s$ with the
probabilities $F_{ij\cdots k}$ and $F_{\bar{i}\bar{j}\cdots
\bar{k}}$, respectively.

Let us first discuss the case that the $N$-photon system is in the
state $\vert \Phi^+ \rangle_s\equiv \vert \phi\rangle_s  \cdot \vert
\Phi^+\rangle_{ij\cdots k}
=\frac{1}{\sqrt{2}}(|a_1\rangle|b_1\rangle \cdots \vert c_1 \rangle
+ |a_2\rangle|b_2\rangle \cdots \vert c_2 \rangle)\cdot
\frac{1}{\sqrt{2}}(|ij\cdots k\rangle + |\bar{i}\bar{j}\cdots
\bar{k}\rangle)_{AB\cdots C}$. After the setup shown in
Fig.\ref{fig2multiparticlespatial}, the evolution of the state
$\vert \Phi^+ \rangle_s$ can be described as

\begin{widetext}
\begin{center}
\begin{eqnarray}
\vert \Phi^+ \rangle_s &\underrightarrow{\; \; HWPs \;\;}&
 \frac{1}{2}(|ij\cdots k\rangle|a_1\rangle|b_1\rangle \cdots \vert c_1 \rangle + |\bar{i}\bar{j}\cdots \bar{k}\rangle|a_2\rangle|b_2\rangle \cdots
\vert c_2 \rangle \nonumber\\
&& \;   +    |\bar{i}\bar{j}\cdots
\bar{k}\rangle|a_1\rangle|b_1\rangle \cdots \vert c_1 \rangle +
|ij\cdots k\rangle|a_2\rangle|b_2\rangle \cdots \vert c_2 \rangle)
\nonumber\\
&\underrightarrow{\; \; PBSs \;\;}&
 \frac{1}{2}(|ij\cdots k\rangle|a'_{\bar{i}}\rangle|b'_{\bar{j}}\rangle \cdots \vert c'_{\bar{k}}
\rangle
 + |\bar{i}\bar{j}\cdots \bar{k}\rangle|a'_{\bar{i}}\rangle|b'_{\bar{j}}\rangle \cdots
\vert c'_{\bar{k}} \rangle  \nonumber\\
&& \;   +    |\bar{i}\bar{j}\cdots \bar{k}\rangle
   |a'_i\rangle|b'_j\rangle \cdots \vert c'_k \rangle + |ij\cdots
k\rangle|a'_i\rangle|b'_j\rangle \cdots \vert c'_k \rangle)
\nonumber\\
&=&  \frac{1}{2} (|ij\cdots k\rangle + |\bar{i}\bar{j}\cdots
\bar{k}\rangle)(|a'_i\rangle|b'_j\rangle \cdots \vert c'_k \rangle +
|a'_{\bar{i}}\rangle|b'_{\bar{j}}\rangle \cdots \vert c'_{\bar{k}}
\rangle)_{AB\cdots C}. 
\end{eqnarray}
\end{center}
\end{widetext}
That is, the parties in quantum communication can determine the
state of their $N$-photon system by postselection. If the photons
emit from the outports  $|a'_i\rangle|b'_j\rangle \cdots \vert c'_k
\rangle$ or $|a'_{\bar{i}}\rangle|b'_{\bar{j}}\rangle \cdots \vert
c'_{\bar{k}} \rangle$ of the PBSs, the system is in the state $
\frac{1}{\sqrt{2}} (|ij\cdots k\rangle + |\bar{i}\bar{j}\cdots
\bar{k}\rangle)$. The parties can obtain the state
$|\Phi^{+}\rangle_N=\frac{1}{\sqrt{2}}(|HH\cdots H\rangle +
|VV\cdots V\rangle)_{AB\cdots C}$ with some unitary single-photon
operations. For example, if the $N$ photons emit from the outports
$|a'_1\rangle|b'_0\rangle \cdots \vert c'_0 \rangle$, the system is
in the state $ \frac{1}{\sqrt{2}} (|10\cdots 0\rangle + |01\cdots
1\rangle)_{AB\cdots C}=\frac{1}{\sqrt{2}} (|V\rangle|H\rangle\cdots
|H\rangle + |H\rangle|V\rangle\cdots |V\rangle)_{AB\cdots C}$ and
the parties can obtain the state $ \frac{1}{\sqrt{2}}
(|H\rangle|H\rangle\cdots |H\rangle + |V\rangle|V\rangle\cdots
|V\rangle)_{AB\cdots C}$ with only a bit-flip operation
$\sigma_x=|V\rangle\langle H\vert + \vert H\rangle\langle V|$ on the
first qubit $A$.

When the $N$-photon system is in the state $\vert
\Phi^-\rangle_s\equiv\vert \phi\rangle_s  \cdot \vert
\Phi^-\rangle_{ij\cdots k}
=\frac{1}{\sqrt{2}}(|a_1\rangle|b_1\rangle \cdots \vert c_1 \rangle
+ |a_2\rangle|b_2\rangle \cdots \vert c_2 \rangle) \cdot
\frac{1}{\sqrt{2}}(|ij\cdots k\rangle - |\bar{i}\bar{j}\cdots
\bar{k}\rangle)_{AB\cdots C}$, one can obtain the similar result.
After the setup shown in Fig.\ref{fig2multiparticlespatial}, the
evolution of the state $\vert \Phi^-\rangle_s$ can be described as
\begin{widetext}
\begin{center}
\begin{eqnarray}
\vert \Phi^- \rangle_{s} &\underrightarrow{\; \; HWPs \;\;}&
 \frac{1}{2}(|ij\cdots k\rangle|a_1\rangle|b_1\rangle \cdots \vert c_1 \rangle + |\bar{i}\bar{j}\cdots \bar{k}\rangle|a_2\rangle|b_2\rangle \cdots
\vert c_2 \rangle \nonumber\\
&& \;   -    |\bar{i}\bar{j}\cdots
\bar{k}\rangle|a_1\rangle|b_1\rangle \cdots \vert c_1 \rangle -
|ij\cdots k\rangle|a_2\rangle|b_2\rangle \cdots \vert c_2 \rangle)
\nonumber\\
&\underrightarrow{\; \; PBSs \;\;}&
 \frac{1}{2}(|ij\cdots k\rangle|a'_{\bar{i}}\rangle|b'_{\bar{j}}\rangle \cdots \vert c'_{\bar{k}}
\rangle
 + |\bar{i}\bar{j}\cdots \bar{k}\rangle|a'_{\bar{i}}\rangle|b'_{\bar{j}}\rangle \cdots
\vert c'_{\bar{k}} \rangle  \nonumber\\
&& \;   -    |\bar{i}\bar{j}\cdots \bar{k}\rangle
   |a'_i\rangle|b'_j\rangle \cdots \vert c'_k \rangle - |ij\cdots
k\rangle|a'_i\rangle|b'_j\rangle \cdots \vert c'_k \rangle)
\nonumber\\
&=&  \frac{1}{2} (|ij\cdots k\rangle + |\bar{i}\bar{j}\cdots
\bar{k}\rangle)(|a'_i\rangle|b'_j\rangle \cdots \vert c'_k \rangle -
|a'_{\bar{i}}\rangle|b'_{\bar{j}}\rangle \cdots \vert c'_{\bar{k}}
\rangle)_{AB\cdots C}. 
\end{eqnarray}
\end{center}
\end{widetext}
The parties in quantum communication can also determine the state of
their $N$-photon system by postselection on the spatial modes of the
$N$ photons. If the photons emit from the outports
$|a'_i\rangle|b'_j\rangle \cdots \vert c'_k \rangle$ or
$|a'_{\bar{i}}\rangle|b'_{\bar{j}}\rangle \cdots \vert c'_{\bar{k}}
\rangle$ of the PBSs, the system is in the state $
\frac{1}{\sqrt{2}} (|ij\cdots k\rangle + |\bar{i}\bar{j}\cdots
\bar{k}\rangle)$. That is, the phase-flip error does not affect the
result and the parties can obtain the state
$|\Phi^{+}\rangle_N=\frac{1}{\sqrt{2}}(|HH\cdots H\rangle +
|VV\cdots V\rangle)_{AB\cdots C}$ with  some unitary single-photon
operations.

\section{Economic one-step MPEECP with frequency entanglement}

Experimental results showed that the frequency of a photon is also
far more stable than its polarization as previous experiments showed
that the polarization entanglement is quite unsuitable for
transmission over distances of more than a few kilometers in an
optical fiber \cite{rmp}. For example, Naik \emph{et al.}
\cite{experiment1} demonstrated the Ekert protocol \cite{Ekert91} by
only a few meters . Also, they observed the quantum bit error rate
(QBER) increase to 33\% in the experiment implementation of the
six-state protocol \cite{sixstate1,sixstate2}. For frequency coding
\cite{experiment2,experiment5,experiment3,experiment4,frequencystable1,frequencystable2},
for example, the Besancon group performed a key distribution over a
20-km single-mode optical fiber spool. They recorded a QBER$_{opt}$
contribution of approximately 4\%, and estimated that 2\% could be
attributed to the transmission of the central frequency by the
Fabry-Perot cavity \cite{experiment5}. That is, the parties in
quantum communication can also exploit the frequency entanglement of
a multipartite entangled system to correct the errors in its
polarization entanglement, similar to the case with a spatial
entanglement. In this way, the parties can obtain the maximally
entangled state $|\Phi^{+}\rangle_N$ with less channels than that
with  spatial entanglement.

An entangled state of an $N$-photon system in both the polarization
and the frequency  degrees of freedom can be written as
\begin{eqnarray}
|\Psi\rangle_s=\frac{1}{2}(|\omega_{1}\rangle|\omega_{1}\rangle
\cdots \vert \omega_{1} \rangle +
|\omega_{2}\rangle|\omega_{2}\rangle\cdots \vert
\omega_{2}\rangle)(|H\rangle|H\rangle\cdots \vert H\rangle
+|V\rangle|V\rangle\cdots \vert V\rangle)_{AB\cdots
C},\label{state3}
\end{eqnarray}
where $|\omega_{1}\rangle$ and $|\omega_{2}\rangle$ are the two
frequency modes for photons. $\vert
\phi\rangle_f=\frac{1}{\sqrt{2}}(|\omega_{1}\rangle|\omega_{1}\rangle
\cdots \vert \omega_{1} \rangle +
|\omega_{2}\rangle|\omega_{2}\rangle\cdots \vert \omega_{2}\rangle)$
represents the frequency entanglement of an $N$-photon system.

The principle of the present economic one-step MPEECP using
frequency entanglement is shown in Fig.
\ref{fig1multiparticlefrequency}. Suppose that the original state of
an $N$-photon system is $\vert \Psi\rangle_s$. After the photons
suffer from the noise in the channels, the state of the system
becomes
\begin{eqnarray}
\rho_s=\rho_p\cdot \rho_f,\label{state4}
\end{eqnarray}
where
\begin{eqnarray}
\rho_f=\vert \phi\rangle_f\langle \phi\vert. \label{state5}
\end{eqnarray}
$\rho_p$ represents the density matrix in the polarization degree of
freedom shown in Eq. (\ref{state62}). The mixed state shown in
Eq.(\ref{state4}) means that the system is in the state $\vert
\Phi^+\rangle_{ij\cdots k}\cdot \vert \phi\rangle_f$ or the state
$\vert \Phi^-\rangle_{ij\cdots k}\cdot \vert \phi\rangle_f$ with the
probabilities $F_{ij\cdots k}$ and $F_{\bar{i}\bar{j}\cdots
\bar{k}}$, respectively.

\begin{widetext}
\begin{center}
\begin{figure}[!h]
\begin{center}
\includegraphics[width=12cm,angle=0]{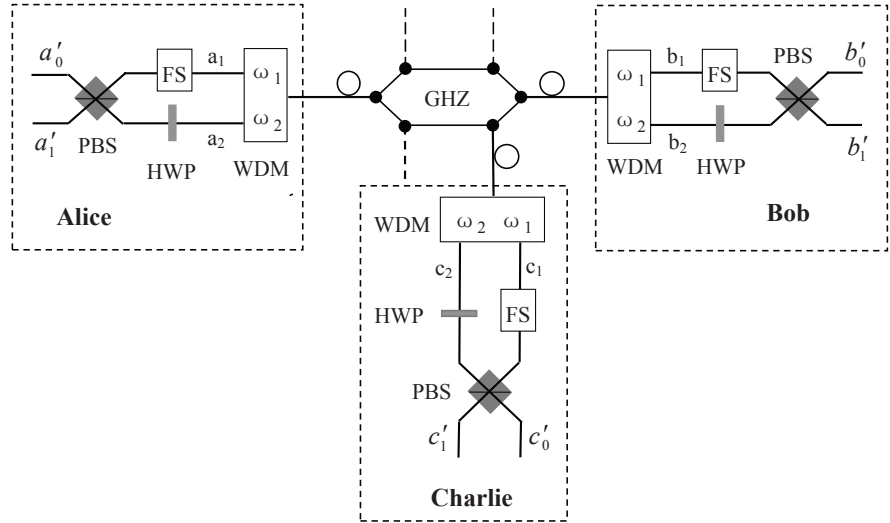}
\caption{Schematic illustration for the principle of the present
economic one-step  multipartite polarization entanglement
error-correction protocol using frequency entanglement. WDM
represents a polarization independent wavelength division
multiplexer which will lead the photons to different spatial modes
according to their frequencies. FS is a frequency shifter which is
used to complete the frequency shift from $\omega_1$ to $\omega_2$.
}\label{fig1multiparticlefrequency}
\end{center}
\end{figure}
\end{center}
\end{widetext}

Similar to the case with the spatial entanglement of an $N$-photon
system, if  the  system is in the state $\vert
\Psi^+\rangle_s\equiv\vert \phi\rangle_f \cdot \vert
\Phi^+\rangle_{ij\cdots k}
=\frac{1}{\sqrt{2}}(|\omega_{1}\rangle|\omega_{1}\rangle \cdots
\vert \omega_{1} \rangle +
|\omega_{2}\rangle|\omega_{2}\rangle\cdots \vert \omega_{2}\rangle)
\cdot \frac{1}{\sqrt{2}}(|ij\cdots k\rangle + |\bar{i}\bar{j}\cdots
\bar{k}\rangle)_{AB\cdots C}$, after the setup shown in
Fig.\ref{fig1multiparticlefrequency}, the evolution of the state
$\vert \Psi^+\rangle_s$ can be described as
\begin{widetext}
\begin{center}
\begin{eqnarray}
\vert \Psi^+ \rangle_{s} &\underrightarrow{\; \; WDMs \;\;}&
 \frac{1}{2}(|ij\cdots k\rangle
+ |\bar{i}\bar{j}\cdots \bar{k}\rangle)_{AB\cdots C} \cdot
(|a_1\rangle_{\omega_{1}}|b_1\rangle_{\omega_{1}} \cdots \vert c_1
\rangle_{\omega_{1}} +
|a_2\rangle_{\omega_{2}}|b_2\rangle_{\omega_{2}} \cdots \vert c_2
\rangle_{\omega_{2}})
 \nonumber\\
&\underrightarrow{\; \; FSs \;\;}&
 \frac{1}{2}(|ij\cdots k\rangle
+ |\bar{i}\bar{j}\cdots \bar{k}\rangle)_{AB\cdots C} \cdot
(|a_1\rangle|b_1\rangle \cdots \vert c_1 \rangle +
|a_2\rangle|b_2\rangle \cdots \vert c_2 \rangle) \nonumber\\
&\underrightarrow{\; \; HWPs \;\;}&
 \frac{1}{2}(|ij\cdots k\rangle|a_1\rangle|b_1\rangle \cdots \vert c_1 \rangle + |\bar{i}\bar{j}\cdots \bar{k}\rangle|a_2\rangle|b_2\rangle \cdots
\vert c_2 \rangle \nonumber\\
&& \;   +    |\bar{i}\bar{j}\cdots
\bar{k}\rangle|a_1\rangle|b_1\rangle \cdots \vert c_1 \rangle +
|ij\cdots k\rangle|a_2\rangle|b_2\rangle \cdots \vert c_2 \rangle)
\nonumber\\
&\underrightarrow{\; \; PBSs \;\;}&
 \frac{1}{2}(|ij\cdots k\rangle|a'_{\bar{i}}\rangle|b'_{\bar{j}}\rangle \cdots \vert c'_{\bar{k}}
\rangle
 + |\bar{i}\bar{j}\cdots \bar{k}\rangle|a'_{\bar{i}}\rangle|b'_{\bar{j}}\rangle \cdots
\vert c'_{\bar{k}} \rangle  \nonumber\\
&& \;   +    |\bar{i}\bar{j}\cdots \bar{k}\rangle
   |a'_i\rangle|b'_j\rangle \cdots \vert c'_k \rangle + |ij\cdots
k\rangle|a'_i\rangle|b'_j\rangle \cdots \vert c'_k \rangle)
\nonumber\\
&=&  \frac{1}{2} (|ij\cdots k\rangle + |\bar{i}\bar{j}\cdots
\bar{k}\rangle)(|a'_i\rangle|b'_j\rangle \cdots \vert c'_k \rangle +
|a'_{\bar{i}}\rangle|b'_{\bar{j}}\rangle \cdots \vert c'_{\bar{k}}
\rangle)_{AB\cdots C}. 
\end{eqnarray}
\end{center}
\end{widetext}
That is, the parties in quantum communication can determine the
state of their $N$-photon system by postselection on the spatial
modes of the photons. If the photons emit from the outports
$|a'_i\rangle|b'_j\rangle \cdots \vert c'_k \rangle$ or
$|a'_{\bar{i}}\rangle|b'_{\bar{j}}\rangle \cdots \vert c'_{\bar{k}}
\rangle$ of the PBSs, the system is in the state $
\frac{1}{\sqrt{2}} (|ij\cdots k\rangle + |\bar{i}\bar{j}\cdots
\bar{k}\rangle)$. The parties can obtain the state
$|\Phi^{+}\rangle_N=\frac{1}{\sqrt{2}}(|HH\cdots H\rangle +
|VV\cdots V\rangle)_{AB\cdots C}$ with some unitary single-photon
operations.

When the $N$-photon system is in the state $\vert
\Psi^-\rangle_s\equiv\vert \phi\rangle_f  \cdot \vert
\Phi^-\rangle_{ij\cdots k}
=\frac{1}{\sqrt{2}}(|\omega_{1}\rangle|\omega_{1}\rangle \cdots
\vert \omega_{1} \rangle +
|\omega_{2}\rangle|\omega_{2}\rangle\cdots \vert \omega_{2}\rangle)
\cdot \frac{1}{\sqrt{2}}(|ij\cdots k\rangle - |\bar{i}\bar{j}\cdots
\bar{k}\rangle)_{AB\cdots C}$, one can obtain the similar result.
After the setup shown in Fig.\ref{fig1multiparticlefrequency}, the
evolution of the state $\vert \Psi^-\rangle_s$ can be described as
\begin{widetext}
\begin{center}
\begin{eqnarray}
\vert \Psi^- \rangle_{s} &\underrightarrow{\; \; WDMs \;\;}&
 \frac{1}{2}(|ij\cdots k\rangle
- |\bar{i}\bar{j}\cdots \bar{k}\rangle)_{AB\cdots C} \cdot
(|a_1\rangle_{\omega_{1}}|b_1\rangle_{\omega_{1}} \cdots \vert c_1
\rangle_{\omega_{1}} +
|a_2\rangle_{\omega_{2}}|b_2\rangle_{\omega_{2}} \cdots \vert c_2
\rangle_{\omega_{2}})
 \nonumber\\
&\underrightarrow{\; \; FSs \;\;}&
 \frac{1}{2}(|ij\cdots k\rangle
- |\bar{i}\bar{j}\cdots \bar{k}\rangle)_{AB\cdots C} \cdot
(|a_1\rangle|b_1\rangle \cdots \vert c_1 \rangle +
|a_2\rangle|b_2\rangle \cdots \vert c_2 \rangle) \nonumber\\
&\underrightarrow{\; \; HWPs \;\;}&
 \frac{1}{2}(|ij\cdots k\rangle|a_1\rangle|b_1\rangle \cdots \vert c_1 \rangle + |\bar{i}\bar{j}\cdots \bar{k}\rangle|a_2\rangle|b_2\rangle \cdots
\vert c_2 \rangle \nonumber\\
&& \;   -    |\bar{i}\bar{j}\cdots
\bar{k}\rangle|a_1\rangle|b_1\rangle \cdots \vert c_1 \rangle -
|ij\cdots k\rangle|a_2\rangle|b_2\rangle \cdots \vert c_2 \rangle)
\nonumber\\
&\underrightarrow{\; \; PBSs \;\;}&
 \frac{1}{2}(|ij\cdots k\rangle|a'_{\bar{i}}\rangle|b'_{\bar{j}}\rangle \cdots \vert c'_{\bar{k}}
\rangle
 + |\bar{i}\bar{j}\cdots \bar{k}\rangle|a'_{\bar{i}}\rangle|b'_{\bar{j}}\rangle \cdots
\vert c'_{\bar{k}} \rangle  \nonumber\\
&& \;   -    |\bar{i}\bar{j}\cdots \bar{k}\rangle
   |a'_i\rangle|b'_j\rangle \cdots \vert c'_k \rangle - |ij\cdots
k\rangle|a'_i\rangle|b'_j\rangle \cdots \vert c'_k \rangle)
\nonumber\\
&=&  \frac{1}{2} (|ij\cdots k\rangle + |\bar{i}\bar{j}\cdots
\bar{k}\rangle)(|a'_i\rangle|b'_j\rangle \cdots \vert c'_k \rangle -
|a'_{\bar{i}}\rangle|b'_{\bar{j}}\rangle \cdots \vert c'_{\bar{k}}
\rangle)_{AB\cdots C}.
\end{eqnarray}
\end{center}
\end{widetext}
That is, the phase-flip error does not affect the result yet, as the
same as the case with spatial entanglement.

\section{discussion and summary}

It is interesting to compare these two economic one-step MPEECPs
with the conventional multipartite entanglement purification
protocols (CMEPPs) in  Refs.\cite{Murao,shengepjd}. In
Ref.\cite{Murao}, Murao \emph{et al.} divided their CMEPP into two
steps. One is used to purify the bit-flip errors and the other is
used to purify the phase-flip errors, by resorting to CNOT gates.
The parties in quantum communication can in principle improve the
fidelity of an ensemble in a multipartite mixed entangled state by
repeating these two steps again and again. The CMEPP in
Ref.\cite{shengepjd} is similar to that by Murao \emph{et al.} but
the CNOT gates are replaced with some parity-check detectors based
on nonlinear optics. That is, the two  existing CMEPPs
\cite{Murao,shengepjd} for $N$-photon systems can be used to improve
the fidelity of an ensemble in a mixed entangled state and they
works in a probabilistic way. In these two economic one-step
MPEECPs, the parties  can in principle obtain a maximally entangled
state from each system transmitted and they works in a completely
deterministic way. On the other hand, the system after it is
transmitted over a noisy channel in the CMEPPs
\cite{Murao,shengepjd} should be in an entangled state in the
polarization degree of freedom. In these two MPEECPs, they do not
require the system to be entangled in the polarization degree of
freedom but they require another degree of freedom of photons to
keep the entanglement.

In essence, these two MPEECPs exploit the entanglement
transformation between two degrees of freedom, not that between two
systems, to correct the errors in the polarization entanglement,
which is different from the CMEPPs \cite{Murao,shengepjd}. In our
two MPEECPs, the polarization degree of freedom of an $N$-photon
system is polluted by the noisy channels. The noise may make the
system be in a completely mixed state in the polarization degree of
freedom. In our first MPEECP, the spatial entanglement is kept as
spatial entanglement is far more stable than the polarization
entanglement over an optical-fiber channel. So does the frequency
entanglement in our second MPEECP. After the transmission over noisy
channels, the parties transform the spatial entanglement into the
polarization entanglement completely in our first MPEECP, by using
some linear optical elements to remove the noise effect on the
polarization first, which is in principle different from the EPP
using spatial entanglement for two-photon systems in Ref.
\cite{Simon}. In our second MPEECP, the parties first exploit WDMs
to transform the frequency entanglement of an $N$-photon system into
its spatial entanglement. With FSs, the parties erase the
distinguishability for the frequencies of their photons. In this
way, the frequency entanglement is completely transformed into the
spatial entanglement. The other precesses are the same as the case
in our first MPEECP. Our second  MPEECP only exploits a fiber
channel for each party, not two channels in our first MPEECP. The
parties can obtain this economization on the number of channels but
they should resort to some nonlinear elements to complete this
MPEECP. This is a tradoff between our two MPEECPs.

In summary, we have presented two  economic one-step  MPEECPs for an
$N$-photon system in a GHZ state. In our first MPEECP, the parties
exploit the spatial entanglement of an $N$-photon system to correct
the errors in its polarization entanglement with some linear optical
elements. Moreover, this protocol works in a deterministic way,
which is far different from the polarization entanglement
purification protocol for two-photon systems by using the spatial
entanglement \cite{Simon} as the latter is used to improve the
fidelity of an ensemble in a mixed entangled state by repeating the
protocol again and again. In our second MPEECP, the parties exploit
the frequency entanglement of an $N$-photon system to correct the
errors in its polarization entanglement, resorting to some nonlinear
optical elements but reducing a fiber channel for each party. As
both the spatial entanglement and the frequency entanglement of an
$N$-photon system are far more stable than the polarization
entanglement over an optical-fiber channel, these two economic
one-step MPEECPs will reduce a greet deal of quantum resource
consumed as they both work in a deterministic way, not a
probabilistic way. This advantage maybe make these two MPEECPs more
useful in practical applications in long-distance quantum
communication in future.



\section*{ACKNOWLEDGEMENTS}

This work is supported by the National Natural Science Foundation of
China under Grant No. 10974020, A Foundation for the Author of
National Excellent Doctoral Dissertation of P. R. China under Grant
No. 200723,  Beijing Natural Science Foundation under Grant No.
1082008, and the Fundamental Research Funds for the Central
Universities.

\end{document}